\documentclass[runningheads]{llncs}
\usepackage[T1]{fontenc}
\usepackage{graphicx}
\usepackage{adjustbox}
\usepackage{orcidlink} 
\usepackage{todonotes}          
\usepackage{enumitem} 
\usepackage{booktabs}
\usepackage{float}
\usepackage{booktabs}
\usepackage{tabularx}
\usepackage{array}
\usepackage{hyperref}
\usepackage{multirow}
\usepackage{ragged2e}
\usepackage{booktabs}
\usepackage{makecell}
\usepackage{xurl}
\usepackage{listings}
\usepackage{xcolor}
\usepackage{longtable}

\begin{document}
\title{Building Knowledge Graphs for Theses: A Survey}
%
%
\author{
Donia Ruelas\inst{1}\orcidlink{0000-0003-1698-1715} \and
Elmer Coyla\inst{1}\orcidlink{0000-0002-9834-4738} \and
Miguel Aceituno\inst{1}\orcidlink{0000-0002-9298-2579} \and
Elwin Huaman\inst{2}\orcidlink{0000-0002-2410-4977}
}
\authorrunning{F. Author et al.}
%
\institute{    
    National University of the Altiplano of Puno, Peru \\
    \email{\{druelas,ecoyla,maceituno\}@unap.edu.pe}
\and    
    University of Cambridge, United Kingdom \\
    \email{elh97@cam.ac.uk}}
\maketitle              
\begin{abstract}
Academic theses are a valuable source of scientific knowledge, yet their semantic exploitation remains limited. While Scholarly Knowledge Graphs have significantly advanced the representation of scientific publications, knowledge graph construction for academic theses has received comparatively limited attention. This paper provides a systematic review of existing approaches and analyzes their potential application for the thesis domain. We analyzed 46 peer-reviewed studies published between 2021 and 2025, and organised existing approaches into a structured classification of techniques, models, and tools. The review reveals a methodological evolution from natural language processing techniques for entity and relation extraction to ontology engineering, Semantic Web technologies, and AI-driven enrichment through knowledge graph embeddings, Linked Open Data, and Large Language Models. The findings provide a consolidated view of the current state of the art and identify methodological foundations and future research directions for building knowledge graphs for the thesis domain.


\end{abstract}
\section{Introduction}
\label{sec:1-intro}

Academic theses constitute one of the largest yet systematically underutilized sources of scientific knowledge within higher education systems~\cite{kousha2020,lariviere2008}. Compared with journal articles, academic theses generally provide more comprehensive descriptions of the research process, including the research problems, objectives, research methodology, empirical analyses, and findings within a single scholarly work. Despite their richness, their visibility and reuse remain limited because institutional repositories primarily rely on keyword-based retrieval and superficial metadata, lacking semantic representations that capture relationships among research entities~\cite{lariviere2008,stocker2023}. Consequently, theses are generally treated as isolated documents rather than interconnected knowledge resources, restricting semantic discovery, longitudinal analysis, and the identification of thematic relationships across research outputs~\cite{hogan2021,tiddi2022}. This limitation is particularly evident in large-scale academic repositories. For example, in Peru, the National University of the Altiplano in the Puno region hosts more than 12,624 theses in its institutional repository\footnote{\url{https://repositorio.unap.edu.pe/}, accessed July 2026.}, while the National Registry of Research Works and Theses (RENATI) consolidates more than 522,334 theses produced by Peruvian universities\footnote{\url{https://renati.sunedu.gob.pe/}, accessed July 2026.}. Despite the availability of such extensive academic collections, the limited semantic organization continues to limit their exploitation for advanced knowledge discovery and scientific analysis.

This situation is also reflected in the current literature on Scholarly Knowledge Graphs. Although substantial progress has been achieved in scholarly knowledge graph construction, existing studies have predominantly focused on journal articles, citation networks, and bibliographic databases~\cite{hogan2021,tiddi2022}. In contrast, academic theses have received comparatively limited attention in the reviewed literature despite their heterogeneous document structures, inconsistent metadata, multilingual content, and domain-specific terminology. Consequently, existing surveys primarily synthesize approaches for scholarly knowledge graphs derived from scientific publications, providing limited discussion on how these methods may be applicable to academic thesis repositories.

To address this gap, this survey systematically reviews recent approaches to scholarly knowledge graph construction and synthesizes the methods and techniques reported in the literature, with particular attention to their potential applicability to academic thesis repositories. The remainder of this paper is organized as follows. Section~\ref{sec:2-methodology} presents the survey methodology; Section~\ref{sec:3-domain} reviews the conceptual foundations of domain-specific knowledge graphs; Section~\ref{sec:4-analysis} synthesizes the reviewed approaches based on the knowledge graph lifecycle identified in the literature.; Section~\ref{sec:5-results} presents the main findings of the literature review; and Section~\ref{sec:6-conclussion} concludes the paper and outlines future research directions.
\section{Survey Methodology}
\label{sec:2-methodology}

This survey follows a systematic literature review methodology inspired by the PRISMA 2020 guidelines~\cite{page2021}. The review protocol was defined prior to the literature search to ensure transparency and reproducibility, encompassing the formulation of research questions, database selection, search strategy, eligibility criteria, study selection, quality assessment, and data extraction.

To guide the review process, this survey addresses the following primary research question: \textit{What approaches exist for building knowledge graphs in the scientific domain?} This question is further decomposed into three sub-research questions, each corresponding to a phase of the knowledge graph lifecycle~\cite{fensel2020,hogan2021} for the thesis domain: 
\begin{itemize}
  \item Which methods and techniques have been proposed for extracting structured knowledge from scientific documents?
  
  \item Which semantic models, ontologies, and graph construction techniques are employed to build knowledge graphs for the scientific domain?
  
  \item Which approaches are used to enrich knowledge graphs in the scientific domain that can improve interoperability, reasoning, and knowledge discovery?
\end{itemize}

\textit{Literature Search.}
A systematic search was conducted in ACM Digital Library, IEEE Xplore, and Scopus using keywords related to scientific knowledge graph construction. Boolean operators were applied and queries were adapted to each database. The search was limited to the 2021–2025.

\textit{Inclusion and exclusion criteria.} 
The inclusion criteria included peer-reviewed journal articles and conference papers addressing scientific knowledge graph construction, published in English between 2021 and 2025, whereas the exclusion criteria involved non-peer-reviewed sources, gray literature, reports, and studies unrelated to knowledge graph construction or semantic representation.

\textit{Data collection.}
After applying the inclusion and exclusion criteria, 296 studies were identified, resulting in 258 unique studies after duplicate removal. A content analysis of titles, keywords, and abstracts was conducted to identify knowledge graph construction approaches, which were subsequently classified according to the knowledge graph lifecycle. Following the screening and quality assessment process, a final corpus of 46 articles was retained for this survey. Subsequently, a comprehensive full-text analysis of the selected articles was conducted to investigate them in detail.

\textit{Categorization in data collection.}
\emergencystretch=3em
The 46 selected papers were classified according to the phases of the knowledge graph lifecycle~\cite{verma2022}: Knowledge Generation, Knowledge Construction, and Knowledge Enrichment~\cite{ji2022,tamasauskaite2022}. These stages were considered because the reviewed studies provided the most comprehensive coverage of methods, techniques, and tools within them. Accordingly, they were used to organize and analyze the approaches to knowledge graph building for the thesis domain.

\section{Domain-Specific Knowledge Graph }
\label{sec:3-domain}
This section explains the concept of a knowledge graph, then defines Thesis Knowledge Graphs and distinguishes them from Scholarly Knowledge Graphs (SKGs) by highlighting their differences.
\paragraph{Knowledge Graph Definition.}
A Knowledge Graph (KG) is a graph-based semantic representation of entities and their relationships, where nodes represent entities and edges encode semantic connections among them~\cite{hogan2021,peng2023}. By integrating heterogeneous information through semantic models, knowledge graphs enable machine-readable representation, interoperability, and querying~\cite{fensel2020}. Owing to these capabilities, KGs have become a fundamental technology for semantic search, knowledge integration, recommendation systems, and scientific knowledge management. Within the scholarly domain, Knowledge Graphs are used to organize scientific knowledge in platforms such as DBpedia, ORKG, Semantic Scholar and Wikidata. 
Table~\ref{tab:kg-platforms} summarizes the key characteristics of these platforms.
\setlength{\intextsep}{5pt}
\begin{table}[ht]
\centering
\scriptsize
\begin{tabularx}{\textwidth}{p{1.5cm}XXXX}
\toprule
\textbf{Attribute} & \textbf{ORKG} & \textbf{Wikidata} & \textbf{DBpedia} & \textbf{Semantic Scholar} \\
\midrule

\renewcommand{\arraystretch}{1.0}

Founded &
\makecell[l]{2018: Sören Auer} &
\makecell[l]{2012: Wikimedia \\  Foundation} &
\makecell[l]{2007: FU Berlin \\  \& Univ. Leipzig} &
\makecell[l]{2015: Allen Institute \\ for AI (AI2)} \\

License &
\makecell[l]{ Open-access,\\community-driven} &
\makecell[l]{Public \\ (CC0 license)} &
\makecell[l]{Public \\ (CC BY-SA 3.0)}&
\makecell[l]{Proprietary \\ Semantic Scholar \\API License Agreement} \\

\makecell[l]{Database \\ System} &
\makecell[l]{ Neo4j \\ (Property Graph\\Database)} &
\makecell[l]{Blazegraph \\ (RDF triplestore)}&
\makecell[l]{Virtuoso \\ (RDF triplestore)} &
\makecell[l]{Proprietary \\ infrastructure}\\

Ontology &
\makecell[l]{ ORKG Ontology\\(RDF/OWL-based)} &
Wikidata Ontology &
\makecell[l]{DBpedia Ontology\\}&
Proprietary semantic schema \\

\makecell[l]{ Entities} &
$\sim$170K resources &
$\sim$122.3M items &
$\sim$6M entities &
$\sim$280M nodes \\

Triples &
$\sim$1.7M &
$\sim$17B+ &
$\sim$8B triples &
N/A \\

Languages &
Primarily English &
Multilingual &
Multilingual &
Primarily English \\

Repository &
\url{orkg.org} &
\url{wikidata.org} &
\url{dbpedia.org} &
\url{semanticscholar.org} \\

\bottomrule
\end{tabularx}
\caption{Comparative overview of ORKG, Wikidata, DBpedia, and Semantic Scholar.}
\label{tab:kg-platforms}
\end{table}

\setlength{\textfloatsep}{2pt}
Although Scholarly Knowledge Graphs (SKGs)~\cite{verma2022} focus on publications, academic theses require richer semantic representations as they capture comprehensive research processes within a single document. For the purposes of this survey, we provide our definition of Knowledge Graphs for the Thesis domain:
\begin{quote}
   \textit{Thesis Knowledge Graph (TKG) is defined as a domain-specific knowledge graph built from academic theses to semantically represent scientific knowledge in a structured, machine-readable, and interoperable format.}
\end{quote}

\begin{table}[ht]
\scriptsize
\begin{tabularx}{\columnwidth}{>{\bfseries}p{1.7cm} p{4.7cm} X}
\toprule
 Dimension & Scholarly Knowledge Graph (SKG) & Thesis Knowledge Graph (TKG)\\
\midrule
\makecell[l]{Primary\\ source} & Scientific articles, conference papers, and other research outputs. &
Thesis documents as the central source of representation. \\

\makecell[l]{Knowledge\\ focus} & 
Abstract, introduction, methodology, result and discussions. &
Research problems, research questions, motivation, objectives, related works, background knowledge, hypothesis, methodologies, results, conclusions and recommendations.\\

Length  & 
Defined by conferences, journals, or editorial teams. &
Defined by university guidelines.\\

Use cases & \makecell[l]{\url{https://graph.openaire.eu/} \\ \url{https://orkg.org/}} &
N/A (to the best our knowledge)\\
\bottomrule
\end{tabularx}
\caption{Comparison between Scholarly Knowledge Graphs (SKGs) and Thesis Knowledge Graphs (TKGs)}
\label{tab:skg_tkg}
\end{table}
Although TKGs and SKGs belong to the academic domain, they differ in scope and representational focus. These differences are summarized in Table~\ref{tab:skg_tkg}.

\section{Knowledge Graph Construction in the Thesis Domain}
\label{sec:4-analysis}
In this section, the reviewed studies are synthesized according to three recurring activities identified across the reviewed literature: Knowledge Generation, Knowledge Construction, and Knowledge Enrichment, which correspond to the stages of a knowledge graph lifecycle~\cite{hogan2021,fensel2020}. These stages are used for organizing existing techniques, methods, and approaches for knowledge graph construction in the thesis domain.

\subsection{Knowledge Generation}
Although primarily built for scholarly knowledge graphs, Knowledge Generation techniques transform unstructured documents into structured semantic representations and could be applied to knowledge graph building for the thesis domain.

\textit{Named Entity Recognition (NER).} 
NER has been widely used to identify and classify relevant entities from unstructured text into predefined semantic categories. If applied to thesis documents, NER could facilitate the extraction of research-related concepts, which could subsequently serve as nodes in a knowledge graph. Recent advances in knowledge graph construction have been largely driven by the integration of Natural Language Processing (NLP) and machine learning techniques for extracting structured knowledge from scientific documents. Existing approaches increasingly combine information extraction, semantic annotation, and ontology-based modeling to identify research entities and relationships suitable for knowledge graph generation~\cite{dessi2022b,dessi2021}. These advances have enabled the automatic construction of large-scale scholarly knowledge graphs, such as Computer Science Knowledge Graph (CS-KG), which represent scientific concepts, research claims, and semantic relationships extracted from academic publications~\cite{dessi2022b}. In addition, semantic technologies and FAIR knowledge infrastructures have improved the interoperability, reuse, and integration of scholarly knowledge across heterogeneous repositories~\cite{hogan2021,stocker2023}. Scholarly NER has evolved toward transformer-based architectures, ontology-aware pipelines, and semantic interoperability. However, existing models remain focused on English scientific articles, highlighting the need to recognize thesis-specific entities in longer and more heterogeneous documents.

\textit{Relation Extraction (RE).}
It identifies semantic relationships among recognized entities, creating typed edges that represent meaningful connections in a knowledge graph. For potential knowledge graphs for the thesis domain, RE techniques could be used to identify semantic relationships among research entities such as research problems, methodologies, datasets, algorithms, and results through semantic predicates. Approaches have evolved from predefined relation schemas and syntactic methods, including dependency parsing and shortest-path analysis~\cite{dessi2022b,dessi2021,xi2025}, toward semantic matching and graph-based learning methods such as RelMatch, Falcon, and Graph Attention Networks, which improve interoperability and contextual understanding~\cite{jaradeh2023,wu2022}. Recent advances further extend RE through link prediction, event-centric extraction, and graph neural network-based dependency analysis to support knowledge graph completion, scientific relation extraction, and dynamic updates~\cite{borrego2025,chang2023,wu2025}. Recent RE approaches have evolved toward semantic matching, graph neural networks, and knowledge graph completion. However, they remain primarily designed for scientific articles, requiring adaptation for document-level relation extraction in thesis repositories.

\textit{Multimodal Information Extraction (MIE).}
This task extends text-based extraction by integrating textual, visual, and structural information to capture knowledge from heterogeneous academic documents. In knowledge graphs for the thesis domain, this is particularly relevant because theses contain figures, tables, mathematical expressions, algorithms, and document structures beyond plain text. Recent approaches combine adaptive extraction pipelines, multimodal representations, and LaTeX-based semantic annotations to support structured knowledge extraction and machine-readable metadata generation~\cite{jaradeh2023,wu2022}. Despite the shift toward multimodal knowledge acquisition, extracting consistent semantic information from domain-specific formats remains a challenge for applying knowledge graph extraction techniques to large-scale thesis domain.

\textit{Entity Linking (EL).}
Aligns extracted entity mentions with canonical entities in knowledge bases or ontologies, ensuring semantic consistency, reducing ambiguity, and promoting interoperability in knowledge graphs for the thesis domain. EL enables the integration of heterogeneous knowledge sources and supports semantic querying and reuse. Recent studies apply EL in federated knowledge graphs and SPARQL generation through entity alignment and mapping~\cite{nguyen2025,Peng2021}, while author disambiguation and entity resolution improve consistency in academic graphs~\cite{chang2023}. Common implementations integrate EL into modular extraction pipelines using tools such as DBpedia Spotlight, TagMe, and EARL~\cite{jaradeh2021}. However, thesis documents remain challenging because domain-specific terminology, multilingual content, and heterogeneous writing styles often produce ambiguous entities.

\textit{Triple Generation and Knowledge Extraction from Text.} Triple Generation refers to techniques that convert extracted entities and relations into subject–predicate–object (S–P–O) triples, which constitute the basic representation of knowledge graphs. Existing approaches generate triples from multiple document levels, including statistical concept co-occurrence~\cite{borrego2025}, Abstract Syntax Tree analysis for software metadata~\cite{haris2023}, extraction of research entities from scientific articles~\cite{Jiomekong2022}, and event-centric methods linking entities and research activities~\cite{chang2023}. Recent studies extend this process toward document-level and ontology-driven extraction, enabling the generation of semantically coherent facts and RDF triples for scholarly knowledge graphs and semantic querying~\cite{brack2022,santamaria2021}. Triple generation is evolving from sentence-level extraction toward document-level semantic representation by leveraging ontology constraints and contextual information. This shift is particularly important for knowledge graphs in the thesis domain, where research contributions are distributed across multiple thesis sections.
\\ \\ 
Knowledge Generation has evolved toward integrated semantic acquisition pipelines based on transformer models, multimodal analysis, ontology-aware extraction, and semantic annotation. However, existing approaches remain focused on scholarly articles, underscoring the need to adapt them to the unique characteristics of thesis repositories. A consolidated overview of representative techniques and application domains in Knowledge Generation is provided in Table~\ref{tab:knowledge_generation}. 
\begin{table}[ht]
\centering
\scriptsize
\renewcommand{\arraystretch}{1.0}

\begin{tabularx}{\textwidth}{p{1.7cm} p{3.5cm} p{2.5cm} X}
\hline
\textbf{Task} &
\textbf{Method} &
\textbf{Application} &
\textbf{Contribution to build TKG} \\
\hline

\multirow{3}{*}{\textbf{NER}}
& SciBERT + BiLSTM~\cite{xi2025}
& Scientific publications
& Identifies research entities for graph nodes. \\

& DyGIE++ + CSO~\cite{dessi2021}
& Computer science
& Detects fine-grained scientific concepts. \\

& Modular NLP pipelines~\cite{jaradeh2023}
& Scholarly repositories
& Scalable entity extraction. \\

\hline

\multirow{3}{*}{\textbf{EL}}
& Federated entity alignment~\cite{Peng2021}
& Distributed KGs
& Aligns equivalent entities across graphs. \\

& Bidirectional mapping~\cite{nguyen2025}
& Question answering
& Supports SPARQL generation. \\

& Canonical entity mapping~\cite{jaradeh2021}
& Scholarly data
& Improves semantic interoperability. \\

\hline

\multirow{4}{*}{\textbf{RE}}
& Ontology-driven schemas~\cite{xi2025}
& Scientific publications
& Defines semantic predicates. \\

& Dependency parsing~\cite{dessi2021,dessi2022b}
& Computer science
& Extracts semantic relations. \\

& Graph Attention Networks~\cite{wu2022}
& Scientific literature
& Learns contextual relations. \\

& Link prediction~\cite{borrego2025}
& Scholarly KGs
& Infers missing relationships. \\

\hline

\multirow{3}{*}{\textbf{\makecell[l]{Triple \\ Extraction}}}
& Co-occurrence analysis~\cite{borrego2025}
& Scientific publications
& Generates semantic triples. \\

& Research contribution extraction~\cite{Jiomekong2022}
& Scholarly literature
& Captures methods, datasets, and findings. \\

& RDF generation~\cite{santamaria2021}
& Institutional repositories
& Produces interoperable RDF triples. \\

\hline

\multirow{2}{*}{\textbf{\makecell[l]{Multimodal\\IE}}}
& Modular IE pipelines~\cite{jaradeh2021}
& Scholarly repositories
& Integrates heterogeneous extractors. \\

& \makecell[l]{PDF/\LaTeX{} analysis~\cite{bless2024,chen2024,Kim2022}}
& Scientific documents
& Extracts visual content and semantic metadata. \\
\hline\textbf{}
\end{tabularx}
\caption{Knowledge Generation candidate methods could support building knowledge graphs for the thesis domain.}
\label{tab:knowledge_generation}
\end{table}

\subsection{Knowledge Construction}
Knowledge Construction comprises the semantic modeling techniques reported in the literature for transforming extracted knowledge into formal graph representations. Within the scope of this survey, these techniques are analyzed as potential building blocks for developing knowledge graphs in the thesis domain.

\textit{Ontology Engineering.}
Provides the semantic foundation of knowledge construction by defining concepts, classes, properties, and constraints that organize thesis knowledge into machine-interpretable structures. In knowledge graphs for thesis domain, ontologies typically model scholarly entities such as Research problems, research questions, motivation, objectives, related works, background knowledge, hypothesis, methodologies, results, conclusions and recommendations.~\cite{Jiomekong2022}. Ontology quality is commonly assessed through criteria including consistency, modularity, completeness, and semantic adequacy~\cite{knoll2022,steinert2025}. Recent approaches increasingly incorporate ontology learning, alignment, and enrichment to automate concept extraction and improve interoperability across heterogeneous knowledge sources~\cite{ansari2025,ashour2022,bless2024,dessi2022b}. The literature shows a shift from manual ontology design toward semi-automatic and knowledge-driven approaches, although thesis heterogeneity and domain-specific terminology remain challenges.

\textit{Semantic Web Standards: RDF, OWL, and SPARQL.}
These standards underpin interoperability in the knowledge graph lifecycle. RDF represents knowledge as subject--predicate--object triples and supports semantic integration~\cite{bless2024}, while OWL defines ontologies and logical constraints for scholarly knowledge representation~\cite{taffa2025b}. However, RDF and OWL alone may not fully capture the complexity of scientific knowledge~\cite{bucur2021}; thus, recent approaches integrate FAIR principles and Linked Data infrastructures to enhance interoperability~\cite{frey2021}. SPARQL enables semantic querying and retrieval, while recent approaches automate competency question and SPARQL query generation to support scholarly knowledge graph evaluation and reuse~\cite{taghzouti2025}. Recent benchmark datasets such as DBLP-QuAD 2.0 further support the evaluation of SPARQL-based question answering over scholarly knowledge graphs by providing complex natural language queries aligned with scholarly entities~\cite{taffa2025b}.
\textit{Schema Design for Knowledge Graphs.}
Specifies node types, relationships, attributes, and structural constraints, defining how knowledge extracted from thesis could be organized and implemented in a graph structure. In scholarly knowledge graphs, schemas commonly represent entities such as abstract, introduction, methodology, result and discussion~\cite{irrera2023}, while advanced approaches use multi-layer structures for static, dynamic, and temporal information~\cite{chang2023}. Platforms such as ORKG employ standardized schemas to support semantic integration and comparative analysis~\cite{knoll2022}. Recent studies also investigate adaptive schema evolution to support continuously updated scholarly knowledge graphs, particularly through XML transformations and Living Literature approaches~\cite{hannah2023,Wijkstra2021}. However, existing methods still struggle to capture the contextual and argumentative complexity of scientific knowledge~\cite{bucur2021}. Table~\ref{tab:knowledge_construction} summarizes representative approaches in the Knowledge Construction stage, showing how ontology engineering, schema design, and Semantic Web standards jointly support the building knowledge graph for the thesis domain.
\begin{table}[ht]
\centering
\scriptsize
\renewcommand{\arraystretch}{1.0}
\begin{tabularx}{\textwidth}{p{1.8cm} p{3.5cm} p{2.5cm} X}
\hline
\textbf{Component} &
\textbf{Representative Approach} &
\textbf{Technologies} &
\textbf{Contribution to build TKG} \\
\hline

\multirow{4}{*}{\textbf{\makecell[l]{Ontology \\ Engineering}}}

& Domain ontology modeling and semantic representation~\cite{Jiomekong2022}
& OWL, Protégé
& Defines the conceptual model of research knowledge. \\

& Ontology evaluation and quality assessment~\cite{knoll2022,steinert2025}
& FAIR, ORKG
& Ensures consistency, modularity, and semantic adequacy. \\

& Ontology alignment across heterogeneous sources~\cite{ansari2025,ashour2022}
& Silk, HL7 FHIR, owl:sameAs
& Improves interoperability among scholarly repositories. \\

& Ontology enrichment and evolution~\cite{bless2024,dessi2022b}
& RDF/XML, DEO
& Extends ontologies with newly extracted knowledge. \\

\hline

\multirow{3}{*}{\textbf{\makecell[l]{Schema \\ Design}}}

& Scholarly graph schemas~\cite{irrera2023,knoll2022}
& Neo4j, ORKG
& Defines graph entities, relations, and structural constraints. \\

& Multi-layer graph modeling~\cite{chang2023}
& Hierarchical graph schema
& Organizes static, dynamic, and temporal knowledge. \\

& Automatic and adaptive schema generation~\cite{bucur2021,hannah2023,Wijkstra2021}
& XML, Linked Data
& Supports schema evolution and richer semantic representation. \\

\hline

\multirow{3}{*}{\textbf{\makecell[l]{Semantic \\  Web \\ Standards}}}

& RDF-based semantic representation~\cite{bless2024}
& RDF, RDF/XML
& Represents research knowledge as interoperable triples. \\

& OWL ontologies and semantic reasoning~\cite{bucur2021,taffa2025b,taghzouti2025}
& OWL, SPARQL
& Enables formal semantics while highlighting current limitations. \\

& FAIR Linked Data infrastructures~\cite{frey2021}
& SHACL, DataID
& Supports semantic integration and interoperable publishing. \\
\hline
\end{tabularx}
\caption{Knowledge Construction candidate approaches could support building knowledge graphs for the thesis domain.}
\label{tab:knowledge_construction}
\end{table}
\\
Knowledge Construction has evolved from manually engineered ontologies toward semi-automatic semantic modeling supported by ontology learning, schema alignment, and FAIR-oriented Semantic Web technologies. Across the reviewed literature, interoperability has become the dominant design objective, motivating the adoption of RDF, OWL, SPARQL, and Linked Data standards. Nevertheless, none of the reviewed studies proposes ontology models specifically designed to represent the semantic structure of academic theses, suggesting that thesis-oriented ontology engineering remains an open research problem.

\subsection{Knowledge Enrichment}
Knowledge Enrichment encompasses the approaches reported in the literature for extending and improving scholarly knowledge graphs. These methods could also support knowledge graphs for thesis domain by improving completeness, interoperability, and semantic reasoning.

\textit{Open Research Knowledge Graph(ORKG)} It is a widely adopted scholarly knowledge graph platform that supports knowledge representation, enrichment, semantic querying, and research management. In the knowledge graph lifecycle, it exemplifies continuous semantic enrichment through automated acquisition and graph-based exploration. Recent work extends ORKG with automated ingestion, alignment with extraction pipelines, and structured representation for literature comparison and systematic review support~\cite{bless2024,brack2022,haris2023,jaradeh2021}. It also enables advanced discovery through SPARQL validation, semantic search, and Large Language Model integration for query generation and question answering~\cite{heidari2021,Jiang2023,meloni2025,muhammad2022,pan2025,pliukhin2023}, ORKG has evolved into a semantic research ecosystem combining knowledge representation and AI-driven services~\cite{Jiomekong2022,kovriguina2024,silva2025}. This evolution is consistent with recent efforts to integrate knowledge graphs into scientific data repositories, enabling richer metadata management, semantic interoperability, and long-term research data reuse. Similar initiatives have extended these ideas to scientific data repositories, where knowledge graphs facilitate semantic metadata management, dataset interoperability, and long-term research data reuse~\cite{pinheiro2024}.

\textit{Encyclopedic Knowledge Graphs and Linked Open Data.}
Encyclopedic Knowledge Graphs and Linked Open Data resources support Knowledge Enrichment by providing external knowledge for entity linking, disambiguation, metadata enrichment, and interoperability. Wikidata and DBpedia are the most widely used resources, while their integration with domain ontologies such as the Computer Science Ontology (CSO) further improves semantic consistency in extraction pipelines~\cite{dessi2022b,frey2021,jaradeh2021}. These resources also underpin SPARQL benchmarking, question answering, and recommendation systems~\cite{Banerjee2023,silva2025,wu2022}. Benchmark datasets such as DBLP-QuAD and DBLP-QuAD 2.0 support the evaluation of natural language question answering through scholarly queries aligned with graph entities~\cite{Banerjee2023,taffa2025b}. Although Linked Open Data substantially improves interoperability and reduces knowledge isolation, thesis-specific concepts often require complementary domain ontologies. It also supports large-scale bibliometric analysis and domain evolution studies~\cite{liu2022}.

\textit{Property Graph:Neo4j.}
Neo4j is a widely used property graph platform for knowledge graphs, providing scalable storage and efficient querying through Cypher. It has been applied to integrate publications, datasets, software, and researchers into unified graphs~\cite{irrera2023}, while ontology-based approaches map semantic classes and properties to support querying and analysis~\cite{li2025}. Neo4j complements RDF with a flexible, high-performance model for knowledge discovery.

\textit{Knowledge Graph Embeddings(KGE)}
Represent entities and relations as low-dimensional vectors, enabling latent semantic pattern learning beyond explicit graph structure. These techniques are widely used for knowledge graph completion tasks, including link prediction, entity alignment, similarity analysis, and graph completion, that could support knowledge graph construction for the thesis domain. Recent work explores scalable and federated embedding models~\cite{Peng2021}, network embedding techniques for large attributed graphs~\cite{yang2022}, and hybrid approaches combining language model embeddings (e.g., MiniLM) with translational models such as TransH, TransD, and RotatE~\cite{borrego2025,chang2023}. More recent studies also investigate hyperbolic embeddings for representing hierarchical academic structures~\cite{rong2025}.

\textit{Large Language Models (LLMs).}
LLMs are key for Knowledge Enrichment, enabling automatic triple generation, knowledge synthesis, query translation, and semantic interaction with graphs. They complement symbolic methods by supporting semantic querying under zero-shot, few-shot, and fine-tuning settings~\cite{Banerjee2023,meloni2025,pan2025}, including query decomposition~\cite{taffa2025a}, structured QA~\cite{Jiang2023}, and multi-agent SPARQL generation~\cite{nguyen2025}. Recent work integrates LLMs with graph retrieval and GraphRAG frameworks for summarization and metadata generation in scholarly knowledge graphs~\cite{chen2024,kovriguina2024}. Recent community efforts have also highlighted Knowledge Graph Construction as an active research area for NLP and LLMs integration, emphasizing scalable extraction and graph generation pipelines~\cite{vakaj2025}. However, limitations such as hallucination and factual inconsistency persist~\cite{borrego2025}, motivating hybrid approaches that combine LLMs with symbolic knowledge graphs for more reliable enrichment. 

Table~\ref{tab:knowledge_enrichment} summarizes the representative platforms, learning techniques, and AI-based approaches employed during the Knowledge Enrichment stage. The comparison highlights their primary capabilities and their contribution to extending, completing, and maintaining knowledge graphs for the thesis domain.

\begin{table}[H]
\centering
\scriptsize
\renewcommand{\arraystretch}{1.0}
\begin{tabularx}{\textwidth}{p{2cm}p{3cm}p{3cm} X}\hline
\textbf{Component} &
\textbf{Representative Platform / Method} &
\textbf{Primary Capability} &
\textbf{Contribution to build TKG} \\
\hline

Knowledge Graph Platform &
\makecell[l]{ORKG~\cite{bless2024,brack2022,haris2023,heidari2021}\\
~\cite{jaradeh2021,Jiang2023,Jiomekong2024,Jiomekong2022,knoll2022,kovriguina2024} \\
~\cite{meloni2025,pliukhin2023,pan2025,silva2025}}
&
Knowledge acquisition, semantic querying, literature comparison &
Supports continuous enrichment, semantic exploration, and research discovery. \\

Linked Open Data &
\makecell[l]{DBpedia, Wikidata \\ ~\cite{Banerjee2023,dessi2022b,frey2021}\\ ~\cite{jaradeh2021,silva2025,wu2022}}&
Entity linking, semantic integration, disambiguation &
Enriches thesis entities with external knowledge and improves interoperability. \\

Property Graph Platform &
Neo4j~\cite{irrera2023,li2025} &
Property graph storage and graph querying &
Provides scalable graph management and semantic navigation. \\

Representation Learning &
Knowledge Graph Embeddings~\cite{borrego2025,chang2023,Peng2021,rong2025,yang2022} &
Link prediction, entity alignment, graph completion &
Discovers implicit relationships and improves graph completeness. \\

Generative AI &
\makecell[l]{Large Language \\ Models~\cite{Banerjee2023,borrego2025,chen2024,Jiang2023}\\
~\cite{kovriguina2024,meloni2025,nguyen2025,pan2025,taffa2025a,vakaj2025}}&
Triple generation, NL-to-SPARQL, semantic QA &
Automates semantic enrichment while requiring mechanisms to mitigate hallucinations. \\
\hline
\end{tabularx}
\caption{Knowledge Enrichment: platforms and methods relevant to build knowledge graphs for thesis domain.}
\label{tab:knowledge_enrichment}
\end{table}
Overall, the reviewed literature shows an evolution in Knowledge Enrichment from platform-centered infrastructures to AI-driven semantic enrichment. Early approaches focused on external knowledge integration using platforms such as ORKG, Wikidata, DBpedia, and Neo4j. Knowledge Enrichment has evolved toward integrating knowledge graphs with Large Language Models to support semantic querying and knowledge completion. However, ensuring factual consistency, provenance, and minimizing hallucinations remains essential for thesis repositories.

\section{Results}
\label{sec:5-results}

This survey reviews advances in scholarly knowledge graph construction and synthesizes the literature for analyzing methods relevant to thesis knowledge graphs through a knowledge graph lifecycle: Knowledge Generation, Knowledge Construction, and Knowledge Enrichment. Together, these three phases encompass semantic extraction, ontology-based graph construction, and AI-driven knowledge enrichment. Despite significant methodological advances, academic theses remain an under-explored source of scientific knowledge.

The analysis of the 46 selected studies reveals a clear methodological evolution in scholarly knowledge graph research. Early approaches primarily focused on information extraction and natural language processing techniques to identify entities, relations, and semantic triples from unstructured scholarly documents. More recent studies increasingly emphasize ontology engineering, schema design, and Semantic Web standards to achieve semantic consistency and interoperability. The latest research extends these approaches through graph embeddings, Linked Open Data resources, and LLMs, enabling more advanced semantic enrichment, knowledge querying, and knowledge discovery.

Across the three lifecycle stages, scholarly knowledge graph research has evolved toward intelligent semantic infrastructures through transformer-based extraction, semantic interoperability, graph learning, and generative AI. However, these approaches remain largely focused on scholarly publications, underscoring the need for adaptation to the unique characteristics of thesis repositories.
\section{Conclusions and Future Work}
\label{sec:6-conclussion}
Academic theses remain one of the largest yet least exploited sources of scientific knowledge because institutional repositories continue to rely primarily on document-oriented organization and limited semantic representation. This survey addresses this gap by systematically synthesizing recent advances in scholarly knowledge graph research through a three-phase knowledge graph lifecycle, such as Knowledge Generation, Knowledge Construction, and Knowledge Enrichment, for analyzing methods relevant to knowledge graphs for the thesis domain.

Knowledge Generation focuses on extracting structured semantic information from unstructured thesis documents, Knowledge Construction formalizes this information through ontology engineering and Semantic Web technologies, and Knowledge Enrichment extends knowledge graphs using external knowledge sources, knowledge graph embeddings, and LLMs to improve graph completeness and semantic querying.

The survey also reveals a clear methodological evolution from traditional information extraction pipelines toward hybrid AI-driven approaches that combine symbolic knowledge representation, graph learning, and generative AI. Nevertheless, important research gaps remain, including the limited focus on academic theses as a distinct source of scientific knowledge, the lack of end-to-end methodologies specifically designed for thesis collections, the challenges posed by heterogeneous thesis structures, and the need for more reliable AI-assisted semantic enrichment.

Future research should focus on scalable, end-to-end methodologies for constructing Thesis Knowledge Graphs using domain-specific ontologies, multilingual information extraction, provenance-aware graph management, and trustworthy LLMs integration. Validation on large institutional and national thesis repositories will be essential to assess their effectiveness for semantic search, research analytics, knowledge reuse, and scientific knowledge discovery, e.g., a thesis knowledge graph for the National University of the Altiplano of Puno.

%
%
%
\bibliographystyle{splncs04}
\bibliography{bibliography}
\clearpage
\section{Appendices}
\label{sec-7:ppendices}

\lstdefinestyle{searchstring}{
    basicstyle=\ttfamily\small,
    breaklines=true,
    breakatwhitespace=true,
    frame=single,
    columns=fullflexible,
    backgroundcolor=\color{gray!5}
}

\appendix

\section{Systematic literature review process used for paper selection}
\label{semantic-literature-review-process}

\begin{figure}[ht]
    \centering
    \adjustbox{angle=90,max width=0.55\textheight,max totalheight=1.1\textwidth}{%
        \includegraphics{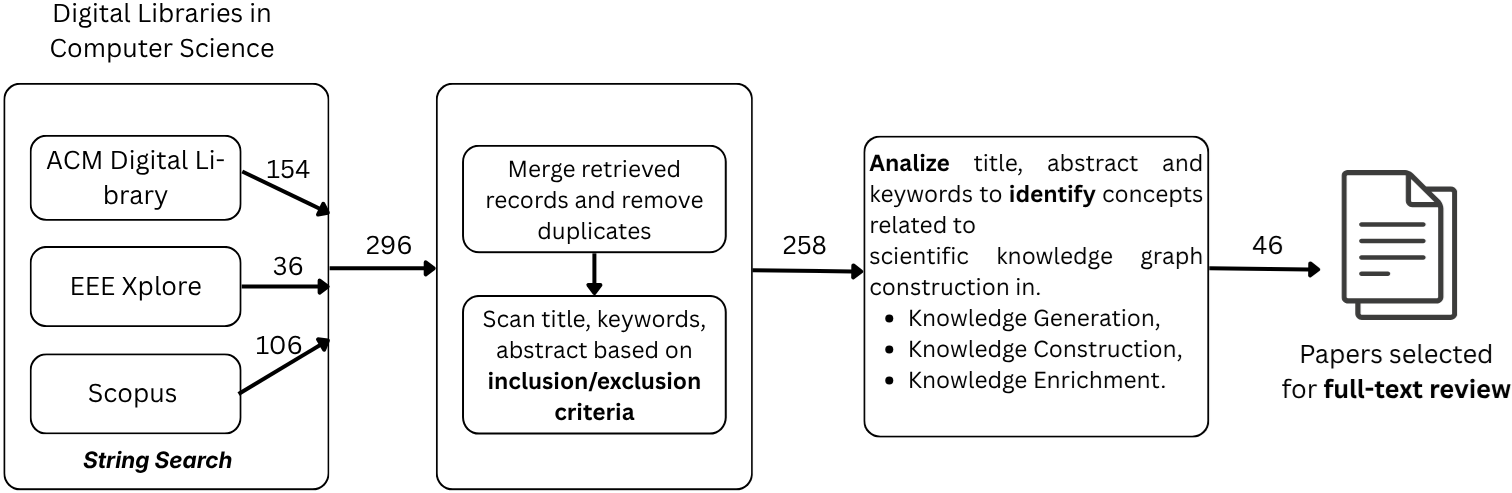}
    }
    \caption{Paper selection process.}
    \label{fig:paper-selection}
\end{figure}

\clearpage
\section{Search String}
\label{app:search-string}

The following search string was used in Scopus:

\begin{lstlisting}[style=searchstring]
TITLE-ABS-KEY ( ( "scientific knowledge graph*" OR "scholarly knowledge graph*"
OR "research knowledge graph*" OR "academic knowledge graph*"
OR "literature knowledge graph*" )
W/15
( construct* OR build* OR built OR develop* OR creat* OR generat* OR design*
OR extract* OR pipeline OR automat* ) )
AND ( LIMIT-TO ( DOCTYPE , "cp" ) OR LIMIT-TO ( DOCTYPE , "ar" ) )
AND ( LIMIT-TO ( LANGUAGE , "English" ) )
AND PUBYEAR > 2020 AND PUBYEAR < 2026
\end{lstlisting}
The following search string was used in ACM Digital Library:

\begin{lstlisting}[style=searchstring]
Search query:
("scientific knowledge graph" OR "scholarly knowledge graph" OR "research knowledge graph" OR "academic knowledge graph" OR "literature knowledge graph")
AND
(construct* OR build* OR develop* OR extract* OR pipeline)
\end{lstlisting}
\noindent\textbf{Filters applied:}
Years 2021--2025 \\\\
The following search string was used in IEEE Explore:
\begin{lstlisting}[style=searchstring]
("All Metadata":"Scientific Knowledge Graph*" OR
"All Metadata":"Scholarly Knowledge Graph*" OR
"All Metadata":"Research Knowledge Graph*" OR
"All Metadata":"Academic Knowledge Graph*" OR
"All Metadata":"literature knowledge graph*")
AND
("All Metadata":"construct*" OR
"All Metadata":"build*" OR
"All Metadata":"develop*" OR
"All Metadata":"generate*" OR
"All Metadata":"create*" OR
"All Metadata":"extract" OR
"All Metadata":"pipeline")
\end{lstlisting}

\noindent\textbf{Filters applied:} Conferences, Journals; Years 2021--2025

\section{List of selected papers}
\label{app:list-selected-paper}
\begin{table}[H]
\centering
\scriptsize
\renewcommand{\arraystretch}{1.0}
\begin{tabularx}{\textwidth}{p{0.5cm} X p{0.5cm} }

\toprule
\textbf{N°} & \textbf{Title} & \textbf{Cit.} \\
\midrule

1 & Multi-Information Preprocessing Event Extraction With BiLSTM-CRF Attention for Academic Knowledge Graph Construction & ~\cite{chang2023}\\

2 & SCIGENT: Multi-agent for End-to-end Question Answering on Scientific Knowledge Graph & ~\cite{nguyen2025}\\

3 & Hyperboloid-Aware Cross-Community Knowledge Graph Contrastive Learning for Paper Recommendation & ~\cite{rong2025}\\

4 & Survey Table Generation from Academic Articles & ~\cite{chen2024}\\

5 & SH-CoDE: Scholarly Hybrid Complex Question Decomposition and Execution & ~\cite{taffa2025a}\\

6 & SciKGTeX -- A LaTeX Package to Semantically Annotate Contributions in Scientific Publications & ~\cite{bless2024}\\

7 & DBLP QuAD 2.0: Scholarly Natural Questions from SPARQL & ~\cite{taffa2025b}\\

8 & 4th International Workshop on Natural Language Processing for Knowledge Graph Construction & ~\cite{vakaj2025}\\

9 & Enhancing Academic Paper Recommendations via Fine-Grained Knowledge Entities and Multidimensional Document Embeddings & ~\cite{xi2025}\\

10 & Demonstration of Faceted Search on Scholarly Knowledge Graphs & ~\cite{heidari2021}\\

11 & Examining the ORKG towards Representation of Control Theoretic Knowledge -- Preliminary Experiences and Conclusions & ~\cite{knoll2022}\\

12 & Q\textsuperscript{2}Forge: Minting Competency Questions and SPARQL Queries for Question-Answering Over Knowledge Graphs & ~\cite{taghzouti2025}\\

13 & A Novel Curated Scholarly Graph Connecting Textual and Data Publications & ~\cite{irrera2023}\\

14 & Exploring Large Language Models for Scientific Question Answering via Natural Language to SPARQL Translation & ~\cite{meloni2025}\\

15 & No PANE, No Gain: Scaling Attributed Network Embedding in a Single Server & ~\cite{yang2022}\\

16 & Designing, Implementing and Deploying an Enterprise Knowledge Graph from A to Z & ~\cite{Jiomekong2022}\\

17 & Differentially Private Federated Knowledge Graphs Embedding & ~\cite{Peng2021}\\

18 & FAIR Linked Data -- Towards a Linked Data Backbone for Users and Machines & ~\cite{frey2021}\\

19 & Bibliometric Analysis of TCM Knowledge Engineering Based on CNKI & ~\cite{liu2022}\\

20 & Choosing the Right Ontology to Describe Research Data in the Energy Domain & ~\cite{steinert2025}\\

21 & Expressing High-Level Scientific Claims with Formal Semantics & ~\cite{bucur2021}\\

22 & Living Literature Reviews & ~\cite{Wijkstra2021}\\

23 & Construction and Leverage Scientific Knowledge Graphs by Means of Semantic Technologies & ~\cite{santamaria2021}\\

24 & A Review of Intelligent Application of Educational Knowledge Graph from the Perspective of Artificial Intelligence & ~\cite{wu2022}\\

25 & A Prototype ETL Pipeline that Uses HL7 FHIR RDF Resources When Deploying Pure Functions to Enrich Knowledge Graph Patient Data & ~\cite{ansari2025}\\

26 & Ontology-Based Linked Data to Support Decision-Making within Universities & ~\cite{ashour2022}\\

27 & Research Knowledge Graphs in NFDI4DataScience: Key Activities, Achievements, and Future Directions & ~\cite{silva2025}\\

28 & Better Call the Plumber: Orchestrating Dynamic Information Extraction Pipelines & ~\cite{jaradeh2021}\\

29 & Research Hypothesis Generation over Scientific Knowledge Graphs & ~\cite{borrego2025}\\

30 & An Approach Based on Open Research Knowledge Graph for Knowledge Acquisition from Scientific Papers & ~\cite{Jiomekong2024}\\

31 & Analysing the Requirements for an Open Research Knowledge Graph & ~\cite{brack2022}\\

32 & SCICERO: A Deep Learning and NLP Approach for Generating Scientific Knowledge Graphs in the Computer Science Domain & ~\cite{dessi2022b}\\

33 & DBLP-QuAD: A Question Answering Dataset over the DBLP Scholarly Knowledge Graph & ~\cite{Banerjee2023}\\

34 & Improving Subgraph Extraction Algorithms for One-Shot SPARQL Query Generation with Large Language Models & ~\cite{pliukhin2023}\\

35 & Generating Knowledge Graphs by Employing Natural Language Processing and Machine Learning Techniques within the Scholarly Domain & ~\cite{dessi2021}\\

36 & A Structure and Content Prompt-based Method for Knowledge Graph Question Answering over Scholarly Data & ~\cite{Jiang2023}\\

37 & Towards a Methodology for the Semi-Automatic Generation of Scientific Knowledge Graphs from XML Documents & ~\cite{hannah2023}\\

38 & Towards Scientific Knowledge Graphs: Dependency Graph Analysis Using Graph Neural Networks for Extracting Scientific Relations & ~\cite{wu2025}\\

39 & Scholarly Knowledge Graph Construction from Published Software Packages & ~\cite{haris2023}\\

40 & Layout Aware Semantic Element Extraction for Sustainable Science \& Technology Decision Support & ~\cite{Kim2022}\\

41 & Information Extraction Pipelines for Knowledge Graphs & ~\cite{jaradeh2023}\\

42 & Knowledge Graph Enabled Scientific Data Repositories & ~\cite{pinheiro2024}\\

43 & FIRESPARQL: A LLM-Based Framework for SPARQL Query Generation over Scholarly Knowledge Graphs & ~\cite{pan2025}\\

44 & Query Resolution of Literature Knowledge Graphs Using Hybrid Document Embeddings & ~\cite{muhammad2022}\\

45 & Aging-Friendly Design Research: Knowledge Graph Construction for Elderly Advantage Applications & ~\cite{li2025}\\

46 & Enhancing Scientific Discovery and Decision-Making: A Knowledge Graph-based Research Support System & ~\cite{kovriguina2024}\\

\bottomrule
\end{tabularx}
\end{table}

\clearpage
\section{Number of Undergraduate Theses at the National University of the Altiplano}
\label{app:number-thesis-unap}

\begin{table}[ht]
\centering
\scriptsize
\renewcommand{\arraystretch}{1.0}
\begin{tabularx}{\textwidth}{>{\centering\arraybackslash}p{3cm} X >{\centering\arraybackslash}p{2cm}}
\hline
\textbf{Area} & \textbf{Academic Program} & \textbf{Number of Theses} \\
\hline

\multirow{6}{*}{\centering \makecell[l]{Biomedical \\ Sciences}}
& Biología & 398\\
& Enfermería & 644\\
& Medicina Humana & 506\\
& Medicina Veterinaria y Zootecnia & 517\\
& Nutrición Humana & 398\\
& Odontología & 391\\
\hline

\multirow{15}{*}{\centering \makecell[l]{Engineering \\ Sciences}}
& Arquitectura y Urbanismo & 310\\
& Físico--Matemática & 73\\
& Ingeniería Agrícola & 410\\
& Ingeniería Agroindustrial & 288\\
& Ingeniería Agronómica & 324\\
& Ingeniería Civil & 429\\
& Ingeniería de Minas & 412\\
& Ingeniería de Sistemas & 213\\
& Ingeniería Electrónica & 202\\
& Ingeniería Estadística e Informática & 259\\
& Ingeniería Geológica & 267\\
& Ingeniería Mecánica Eléctrica & 311\\
& Ingeniería Metalúrgica & 155\\
& Ingeniería Química & 188\\
& Ingeniería Topográfica y Agrimensura & 154\\
\hline

\multirow{4}{*}{\centering \makecell[l]{Economic and \\ Business y \\ Sciences}}
& Ciencias Contables & 880\\
& Administración & 458\\
& Ingeniería Económica & 537\\
& Turismo & 252\\
\hline

\multirow{10}{*}{\centering \makecell[l]{Social\\ Sciences}}
& Antropología & 200\\
& Arte & 99\\
& Ciencias de la Comunicación Social & 243\\
& Derecho & 423\\
& Educación Física & 316\\
& Educación Inicial & 361\\
& Educación Primaria & 313\\
& Educación Secundaria & 926\\
& Sociología & 266\\
& Trabajo Social & 501\\
\hline
\multicolumn{2}{c}{\textbf{Total}} &
\textbf{12,624}\\
\hline
\end{tabularx}
\label{tab:tesis_unap}
\end{table}

\section{Number of Undergraduate Theses at Peruvian Universities}

\scriptsize
\begin{longtable}{p{2cm}p{8cm}r}
\label{tab:peruvian_universities}\\
\toprule
\textbf{Type} & \textbf{University} & \textbf{Count}\\
\midrule
\endfirsthead
\toprule
\textbf{Type} & \textbf{University} & \textbf{Count}\\
\midrule
\endhead
\multirow{38}{*}{\textbf{Public}} & Universidad Nacional Agraria La Molina & 2,322\\
& Universidad Nacional Agraria de la Selva & 434\\
& Universidad Nacional Amazónica de Madre de Dios & 1,094\\
& Universidad Nacional Autónoma de Chota & 12\\
& Universidad Nacional Autónoma de Tayacaja Daniel Hernández Morillo & 24\\
& Universidad Nacional Federico Villarreal & 10,495\\
& Universidad Nacional Hermilio Valdizan & 7,830\\
& Universidad Nacional Intercultural de la Amazonía & 177\\
& Universidad Nacional Intercultural de la Selva Central Juan Santos Atahualpa & 1\\
& Universidad Nacional José Faustino Sánchez Carrión & 5,797\\
& Universidad Nacional José María Arguedas & 61\\
& Universidad Nacional Mayor de San Marcos & 27,533\\
& Universidad Nacional Micaela Bastidas de Apurímac & 1,392\\
& Universidad Nacional San Luis Gonzaga & 4,070\\
& Universidad Nacional Santiago Antúnez de Mayolo & 339\\
& Universidad Nacional Tecnológica de Lima Sur & 1,387\\
& Universidad Nacional Toribio Rodríguez de Mendoza & 1,637\\
& Universidad Nacional de Barranca & 336\\
& Universidad Nacional de Cajamarca & 124\\
& Universidad Nacional de Cañete & 4\\
& Universidad Nacional de Educación Enrique Guzmán y Valle & 1,025\\
& Universidad Nacional de Frontera & 1\\
& Universidad Nacional de Huancavelica & 4,606\\
& Universidad Nacional de Ingeniería & 10,445\\
& Universidad Nacional de Jaén & 341\\
& Universidad Nacional de Moquegua & 55\\
& Universidad Nacional de Música & 56\\
& Universidad Nacional de Piura & 79\\
& Universidad Nacional de San Agustín & 9,492\\
& Universidad Nacional de San Cristóbal de Huamanga & 1,403\\
& Universidad Nacional de San Martin -- Tarapoto & 4,572\\
& Universidad Nacional de Trujillo & 8,379\\
& Universidad Nacional de Tumbes & 1,246\\
& Universidad Nacional de la Amazonía Peruana & 5,490\\
& Universidad Nacional del Altiplano & 14,723\\
& Universidad Nacional del Callao & 5,534\\
& Universidad Nacional del Centro del Perú & 5,578\\
& Universidad Nacional del Santa & 829\\
\midrule
\multirow{53}{*}{\textbf{Private}} & Pontificia Universidad Católica del Perú & 24,086\\
& Universidad Alas Peruanas & 11,290\\
& Universidad Andina del Cusco & 9,075\\
& Universidad Antonio Ruiz de Montoya & 539\\
& Universidad Autónoma de Ica & 5\\
& Universidad Autónoma del Perú & 3,167\\
& Universidad Católica San Pablo & 1,299\\
& Universidad Católica Santo Toribio de Mogrovejo & 6,683\\
& Universidad Católica Sedes Sapientiae & 2,408\\
& Universidad Católica de Santa María & 4,559\\
& Universidad Católica de Trujillo Benedicto XVI & 4,240\\
& Universidad Católica los Ángeles de Chimbote & 2,198\\
& Universidad Científica del Perú & 2,415\\
& Universidad Científica del Sur & 2,173\\
& Universidad Continental & 8,043\\
& Universidad César Vallejo & 165,562\\
& Universidad ESAN & 445\\
& Universidad Femenina del Sagrado Corazón & 813\\
& Universidad Inca Garcilaso de la Vega & 1,167\\
& Universidad Jaime Bausate y Meza & 107\\
& Universidad José Carlos Mariátegui & 19\\
& Universidad La Salle & 14\\
& Universidad Le Cordon Bleu & 6\\
& Universidad Marcelino Champagnat & 442\\
& Universidad María Auxiliadora & 2,302\\
& Universidad Peruana Cayetano Heredia & 3,184\\
& Universidad Peruana Los Andes & 8,861\\
& Universidad Peruana Unión & 7,119\\
& Universidad Peruana de Ciencias Aplicadas & 7,585\\
& Universidad Peruana de Ciencias e Informática & 72\\
& Universidad Peruana del Centro & 5\\
& Universidad Privada Antenor Orrego & 11,140\\
& Universidad Privada Norbert Wiener & 1,292\\
& Universidad Privada San Juan Bautista S.A.C & 5,591\\
& Universidad Privada Telesup & 88\\
& Universidad Privada de Huancayo Franklin Roosevelt & 159\\
& Universidad Privada de Tacna & 3\\
& Universidad Privada del Norte & 20,161\\
& Universidad Ricardo Palma & 7,055\\
& Universidad San Ignacio de Loyola & 4,714\\
& Universidad San Pedro & 1,105\\
& Universidad Señor de Sipán & 11,078\\
& Universidad Tecnológica de los Andes & 855\\
& Universidad Tecnológica del Perú & 13,752\\
& Universidad de Ciencias y Artes de América Latina & 104\\
& Universidad de Ciencias y Humanidades & 6\\
& Universidad de Huánuco & 2,654\\
& Universidad de Ingeniería y Tecnología & 247\\
& Universidad de Lima & 5,373\\
& Universidad de Piura & 1,575\\
& Universidad de San Martín de Porres & 13,693\\
& Universidad del Pacífico & 2,843\\
& Universidad para el Desarrollo Andino & 40\\
\midrule
\textbf{Total} && \textbf{522,334}\\
\bottomrule
\end{longtable}
\footnotesize

\end{document}